\documentclass[prb,aps,twocolumn,showpacs,amsmath,amssymb]{revtex4-1}
\usepackage{graphicx}
\usepackage{dcolumn}
\usepackage{color}
\usepackage{epstopdf}

\begin{document}

\title {High Efficiency CVD Graphene-lead (Pb) Cooper Pair Splitter}

\author {I. V. Borzenets,$^{1}$,Y. Shimazaki$^{1}$, G. F. Jones,$^{2}$ M. F. Craciun,$^{2}$ S. Russo,$^{2}$ Y. Yamamoto,$^{1}$$^{3}$ and S. Tarucha$^{1}$$^{4}$}

\affiliation{
$^{1}$Department of Applied Physics, University of Tokyo, Bunkyo-ku, Tokyo 113-8656, Japan
$^{2}$Centre for Graphene Science, University of Exeter, Exeter EX4 4QL , UK
$^{3}$PRESTO, JST, Kawaguchi-shi, Saitama 332-0012, Japan
$^{4}$Center for Emergent Matter Science (CEMS), RIKEN, Wako-shi, Saitama 351-0198, Japan
 }

\begin{abstract}

We demonstrate high efficiency Cooper pair splitting in a graphene-based device.  We utilize a true Y-shape design effectively placing the splitting channels closer together: graphene is used as the central superconducting electrode as well as QD output channels, unlike previous designs where a conventional superconductor was used with tunnel barriers to the quantum dots (QD) of a different material. Superconductivity in graphene is induced via the proximity effect, thus resulting in both a large measured superconducting gap $\Delta=0.5$meV, and a long coherence length $\xi=200$nm. The graphene-graphene, flat, two dimensional, superconductor-QD interface lowers the capacitance of the quantum dots, thus increasing the charging energy $E_C$ (in contrast to previous devices). As a result we measure a visibility of up to 96\% and a splitting efficiency of up to 62\%. Finally, the devices utilize graphene grown by chemical vapor deposition allowing for a standardized device design with potential for increased complexity. 

\end{abstract}


\maketitle

A source of quantum entangled particles is essential to quantum information processing\cite{1,2}.  The generation of entangled photons has been achieved quite a long time ago\cite{3}. However, the demonstration of a solid state entangler device able to source reproducibly entangled pairs of electrons in an electronic circuit on a chip has only recently been achieved\cite{4,5,6}.  According to BCS theory, electrons in a superconductor naturally form entangled spin singlets known as Cooper pairs\cite{7,8}. A device that can spatially separate the entangled electrons in a superconductor into two normal leads is known as a Cooper Pair splitter\cite{9,10}. High efficiency Cooper Pair splitting (CPS) devices have been made using the “superconductor-two quantum dot” design\cite{10,11}. Such devices have been made using one dimensional nanowires or nanotubes with the central superconductor of Al in a T-shape \cite{11,12,13,14,15}, with reported efficiencies of up to 90\%\cite{14}. Never the less, the one dimensional nature of the nanowire design limits further complexity of such devices. 

Graphene is a single layer, crystalline sheet of carbon with hexagonal structure leading to a gamut of unique electronic properties which would contribute well to a CPS device\cite{16,17, CB}. Indeed, Cooper pair splitting in graphene quantum dots (QD) coupled to a conventional-narrow superconducting wire has recently been demonstrated \cite{18}. In this T-shape design the spatially separated  QDs are partially covered by the superconducting Al wire, resulting in a small charging energy of just 80 $\mu$eV which affects adversely the CPS efficiency of the device - only an efficiency up to 10\% has been reported. In this work we utilize the two dimensional and highly crystalline nature of graphene to achieve a high CPS efficiency device, up to 62\%.  

The ratio between the current due to Cooper pair splitting $(\Delta I)$ to the background current due to two electron processes $(I_{BG})$ is  $\frac{\Delta I}{I_{BG}} =\alpha \frac{2\epsilon^2}{\gamma^2}$ \cite{10}. The prefactor $\alpha$ depends exponentially on the superconducting coherence length $\xi$ and the separation of the normal-metal output channels $\Delta r$ according to the relation: $\alpha=(\sin{(⁡k_F \Delta r)}/(k_F \Delta r) )^2  \exp{(-2\Delta r/\pi\xi)}$. The prefactor $\epsilon$ is a function of the superconducting gap $\Delta$ and the quantum dot charging energy $E_C$: $\frac{1}{\epsilon}=\frac{1}{\pi\Delta}+\frac{1}{E_C}$ . Finally, $\gamma$ is the level resonance width which varies with the gate voltages. We tune the design parameters to maximize $\alpha$ and $\epsilon$.  In contrast to previous works in which Cooper Pairs tunnel from the superconducting metal directly into the QD, our experiments exploit three fundamental elements of novelty which increase dramatically the Cooper pair splitting efficiency. We induce superconductivity in bulk graphene via the proximity effect prior to splitting. Using graphene as the superconductor allows us to increase $\xi$ while keeping $\Delta$ large. Moreover, the device is patterned into a true Y-shape (Figs 1(a), 1(b)),  placing the output channels maximally close together, minimizing $\Delta r$. Finally, the flat, two dimensional nature of the superconductor-quantum dot interface greatly lowers the capacitance of the quantum dots (compared to the device where the superconductor overlays the quantum dot) resulting in a large $E_C$. In our Y-shape Cooper pair splitter we achieve a 100 \% larger value for $\alpha$ and a full order of magnitude larger $\epsilon$ than previously demonstrated in a T-shape QD geometry with graphene \cite{18, note1}. Maximizing these prefactors allows us to find a gate region where the splitting visibility is up to 96\% and an efficiency of 62\% that is 6 times more efficient than previously demonstrated.

\begin{figure}[]
\includegraphics[width=1 \columnwidth]{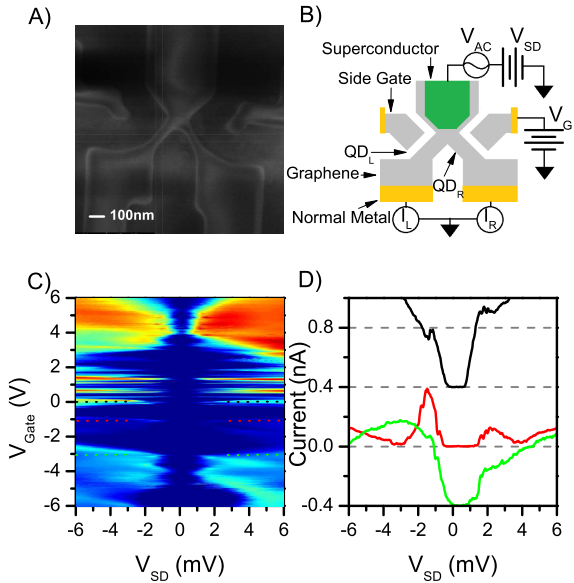}
\caption{\label{fig:overview} a)Scanning Electron micrograph of the active area of the device prior to metal deposition. b)Device and measurement schematic. CVD graphene is patterned into a “Y” shape in order to minimize separation between the normal metal channels. Entrances to the Right and Left normal channels are constricted into nanoribbons. Shown in Fig 1(a) is the nanoribbon of $70$nm width and $\sim150$nm length. The nanoribbons form quantum dots ($QD_R$, $QD_L$) due to edge defects. The central electrode acts as the superconducting lead, Pb is deposited over the central electrode to induce superconductivity. Measurements are done via the lock-in method. AC voltage $V_{AC}$ with a possible bias $V_{DC}$ is applied to the central lead, currents though the Right ($I_R$) and Left($I_L$) channels are measured simultaneously. $QD_R$ and $QD_L$ are controlled in tandem via the back gate, or individually via the self-aligned side gates. c)AC current versus the bias voltage $V_{SD}$ and back gate $V_{BG}$ though one of the channels at base temperature and zero magnetic field. Coulomb blockade peaks separated by a gap can be seen. The presence of the peak shows that quantum dots indeed form, and a measured charging energy of $\sim5$meV if much larger than the superconducting gap of Pb. d) Current v.s. bias voltage $V_{SD}$ taken at the apexes of three Coulomb peaks as marked in Fig 1(c). (The data is offset by $0.4$nA). A superconducting gap is clearly present with a maximum value of $1$meV. 
}
\end{figure}

In a step further, we move away from the conventional method of mechanically exfoliated graphene as it has many of the same drawbacks as the nanowire based samples. Such as: the unpredictable size and location of the graphene crystals requiring that each device must be designed and aligned individually, and the small average size of the exfoliated crystals which places limits on the device complexity. Recent advances in growing macroscopic sized sheets of high quality graphene using Chemical Vapor Deposition (CVD)  \cite{SR} allow for more controlled sample design creating regular arrays of standardized devices \cite{19,20,21,22}. Hence, we utilize high quality CVD graphene as the material of choice in our devices, paving the way towards integrating the Cooper pair splitter devices into more complex circuitry. Graphene is grown on Copper via Chemical vapor deposition method with (minimal defects) and domain sizes of $\sim 100\mu$m, which is much larger than the critical area of the device of a few $\mu$m \cite{SR}. The CVD graphene is cut into squares of approximately $5$mm per side and transferred on to a $297$nm oxide $Si/SiO_2$ wafer using a standard $FeCl_3$, PMMA transfer method \cite{19,20}. An array of $12/150$nm Cr/Au bonding pads which includes an alignment pattern is deposited on the wafer using electron beam lithography (EBL). As a second step, the active area of the graphene is defined with EBL and unnecessary material is etched away using Argon Plasma. This defines the central superconducting lead, the quantum dot constrictions with self-aligned gates, and the long (much greater than the proximity effect) normal metal leads. Finally, the superconducting contact is deposited by electron beam evaporation (Pd/Pb with 6/120nm thickness)  \cite{23,24}.
 
The sample is patterned into a “Y-shaped” junction to minimize separation between the normal metal channels (Figs 1(a),1(b)). The distance between the outer edges of the channels at the point where they contact the central superconducting lead, $\Delta r$, is 140nm. The central lead acts as the superconductor via the proximity effect by having the superconductor placed close to, but not touching the exit channels. Quantum dot constrictions at the entrance of the normal metal leads are created by patterning graphene into regions 70nm wide and 100-150nm long  \cite{25,26}. The two quantum dot channels are individually controlled by graphene self-aligned side-gates. Measurements are taken in a dilution refrigerator at the base temperature of 100mK using lock-in technique. An AC voltage of ($350\mu$V) and a frequency of 447Hz is applied to the central lead, with a possible DC bias offset. Currents through the right and the left channels are measured simultaneously. The channels can be controlled individually via the side-gates, or simultaneously via the heavily doped substrate which acts as a global back-gate.

Initial characterization measurements are presented in Figs 1(c),1(d).  A map of the AC current through one of the channels is presented versus bias voltage and back gate. Coulomb blockade peaks separated by a zero-bias, superconducting gap can be seen. Due to the fact that quantum dots in graphene constrictions on $SiO_2/Si$ are formed by defects, the Coulomb peaks are irregularly spaced. However, we estimate that the quantum dot charging energy $E_C$ is around 5meV, which is greater than the energy gap of Pb (1.2meV \cite{8,27}). (Since quantum dot levels in graphene are four fold degenerate, therefore a large level spacing is not sufficient, and $E_C \gg \Delta$ is required.)   Having the superconducting contact in the plane of the quantum dot (as opposed to over it \cite{18}), reduces the quantum dot capacitance, increasing $E_C$. Several cuts of the current versus the bias voltage are taken at values of the back gate voltages that correspond to the apexes of the Coulomb peaks, shown in Fig 1(d). A clear superconducting gap $\Delta$ can be seen with the energy of 0.5-1.0meV. This value is suppressed compared to the accepted value for pure Pb, probably due the presence of Pd sticking layer.  The measured value of $\Delta$ allows us to calculate a lower bound for the superconducting coherence length $\xi$. For clean graphene, the BCS superconducting coherence length is $\xi_0=\hbar v_F / \Delta=0.66-1.3\mu$m, with $v_F=10^6$m/s being the Fermi velocity and a maximum $\Delta=0.5-1$meV \cite{28}. In a diffusive superconductor the length is also a function of the mean free path $l$ with $\xi= (\xi_0 l)^{0.5}$\cite{29} . We calculated the lower bound for the mean free path $l=60$nm, giving us a minimal coherence length of $\xi=199$nm.
 
Grounding the back gate for stability, we now individually control the quantum dots associated with the Left ($QD_L$) and Right ($QD_R$) channels via the side gates. Figs 2(a),2(b) shows a map of the current $I_L$ though $QD_L$ (Fig 2(a)) and current $I_R$ though $QD_R$ ( Fig 2(b)) versus the Left and Right side gate voltages ($V_{GL}$,$V_{GR}$). Several Coulomb blockade resonances can be seen for both quantum dots. The difference in efficiencies for a given quantum dot of the side gates (as well as the back gate) scales approximately with the distance of the gate to the channel constriction \cite{30}. The cross-talk between $V_{GL}$ and $V_{GR}$ results in a greater variability of the quantum dot tunneling barriers increasing the tunability of $\gamma$.

\begin{figure}[]
\includegraphics[width=0.8 \columnwidth]{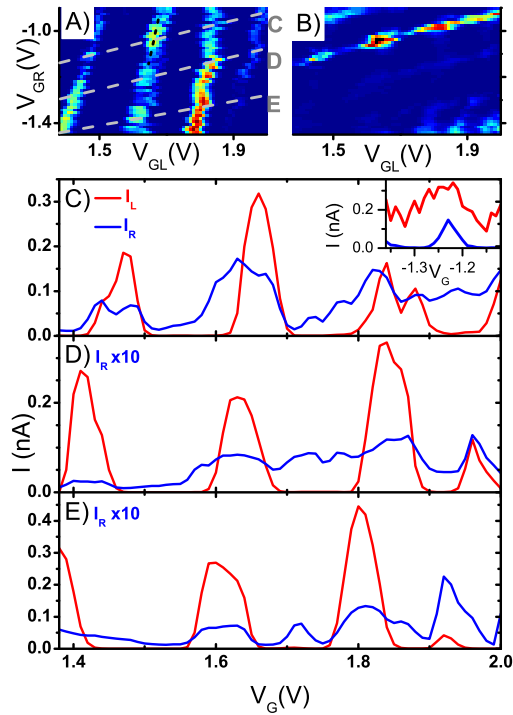}
\caption{\label{fig:overview}a,b)Currents  $I_L$(a) and $I_R$(b) versus the Left $V_{GL}$ and Right $V_{GR}$ side gates. Several Coulomb blockade peaks can be seen for each channel. Visually, it can be seen that the current through the channel is enhanced when both quantum dots are on resonance.  c,d,e)Current versus convolved gate voltage $V_G$, though $QD_R$ (blue) and $QD_L$(red) . The gate voltages are chosen such that $QD_R$ is always kept on resonance $QD_L$ is swept through several conductance peaks. Graphs C,D,E are taken for the three different resonance peaks of $QD_R$ and are represented visually by the dashed lines in Fig 2(a). Clearly, $I_R$ is increased when $QD_L$ goes on resonance: contrary to the classical picture, and a signature of Cooper pair splitting.  C inset) Current versus convolved gate voltage $V_G$, through $QD_R$(blue) and $QD_L$(red). $V_G$ is chosen such that $QD_L$ is kept on resonance while $QD_R$ is swept though a conductance peak: represented by the black dashed line in Fig2(a).
}
\end{figure}

We now look at how $I_R$ evolves as $QD_L$ is tuned on and off resonance\cite{11}. For this, we look at the data with the side gate voltages tuned such that $QD_R$ always remains on resonance; i.e., the data taken along the three dashed lines in Fig2a corresponding to the three resonance peaks of $QD_R$. The $I_L$ and $I_R$ vs gate voltage $V_G$ while keeping $QD_R$ on resonance is presented in Figs 2(c),2(d),2(e).  One can see that since $QD_R$ is kept on resonance, it always has a non-zero conductance. However, $QD_L$ goes through several resonance peaks as the gate voltage is changed. When $QD_L$ is tuned to it’s own Coulomb resonance, a clear enhancement of $I_R$ can be seen. This is taken as a signature of Cooper pair splitting, as having both quantum dots on resonance allows the electrons in a Cooper pair to leave the superconductor efficiently though the left and right dot\cite{10,11}. For other gate configurations CPS is suppressed since the quantum dot constrictions do not allow two-electron tunneling processes. Moreover, enhanced conductance when both channels are on resonance is contrary to a classical picture of a biased three resistor “Y-junction” where the current though one channel would decrease as the conductivity of the other channel increases. 

For comparison, the Fig 2(c) inset shows the change of $I_L$ as $QD_R$ is swept through a resonance. (The data shown in Fig 2(c) inset is taken along the black dashed line presented in Fig 2(a)). As with the case for $QD_R$, $I_L$ is enhanced when both channels are on resonance. However, $QD_L$ has a much higher background current most likely due to much more open tunnel barriers of the quantum dot constriction.  We calculate the current due to Cooper pair splitting by subtracting the background current from the peak current. The background is calculated by averaging $I_L$($I_R$) when $QD_R$ ($QD_L$) is moved directly off-resonance, that is $\Delta I=I_{Peak}-I_{BG}$ which gives  $\Delta I_R=0.11$nA and $\Delta I_L=0.17$nA. Hence we find that the CPS current does not have the same value for both channels. We attribute this imbalance to the finite interdot coupling present in our device which makes it possible for the split electrons to tunnel through the graphene from the Right to the Left quantum dot \cite{12,14,15}. In our device geometry the interdot coupling in mainly present due to the close proximity of the quantum dot entrance constrictions and the presence of a continuous graphene crystal which connects the left and the right hand side of the Y-shape devices. Moreover, $QD_L$ featured much weaker tunneling barriers than $QD_R$ as evidenced by the higher background current; meaning, the time the electron would spend in $QD_L$ would be much shorter than in $QD_R$.  

We calculate the visibility of the Cooper pair splitting by finding the fraction of the CPS current relative to the total current: $\eta=\Delta I/(\Delta I+I_{BG})$. For the resonance peak presented in Fig 2(c) and inset we find $\eta_{Right}=0.92$ and $\eta_{Left}=0.51$. The lower visibility of $QD_L$ is due to the high background current.  Finally we calculate the splitting efficiency of our system by comparing the CPS current to the total current in the system: $s=(\Delta I_{Right}+\Delta I_{Left})/(I_{PeakLeft}+I_{PeakRight})$. We find the efficiency $s=0.62$. This is significantly higher than previously reported in graphene, but much lower than needed to observe a violation of Bell’s inequality\cite{31,32}.

\begin{figure}[]
\includegraphics[width=0.8 \columnwidth]{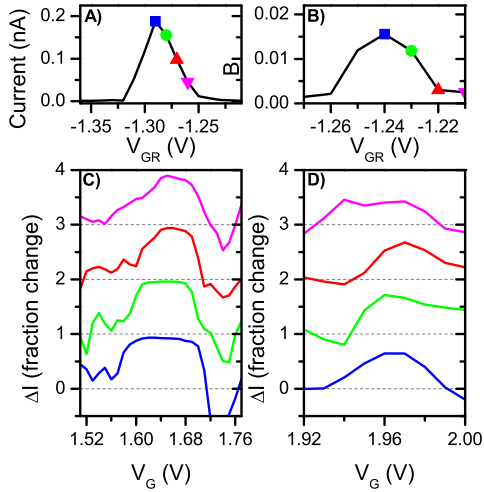}
\caption{\label{fig:overview} a,b) $I_R$ versus the Right gate $V_{GR}$. The gate voltage is swept such that $QD_R$ goes through a resonance peak.  The peak in Fig 3(a) corresponds to the peak presented in Fig 2(c), while the peak in Fig 3(b) corresponds to the peak presented in Fig 2(d). c,d)Fraction change of $I_R$ versus the local background current as $QD_L$ is swept through a resonance. The different curves correspond to different positions away from the resonance as presented in Figs 3(a),3(b) respectively. The background is taken as the average of several data points immediately after $QD_L$ goes off resonance.  The peak current (and therefore peak Cooper pair splitting efficiency) relative to the background happens when $QD_R$ is kept slightly off resonance ($96\%$ v.s.$92\%$ on resonance Fig 3(c); and $71\%$ v.s. $65\%$ for Fig 3d). This is in agreement with previous works. Moreover, the splitting efficiency remains high all through the resonance peak, only falling a factor of $1.6$ at the edge of the resonance. 
}
\end{figure}

Previous works showed that the CPS efficiency was not maximum directly at the top of the quantum dot Coulomb resonance, but instead slightly off peak\cite{10,14}. This was attributed to the fact that on resonance the number of electrons in a quantum dot is not well defined\cite{10}. Therefore, we look at how the relative magnitude of the splitting signature evolves as a function of the off resonance gate shift applied to $QD_R$ while sweeping the Left channel gate. Figs 3(a) and 3(b) shows cuts with respect to the Right gate ($V_{GR}$) for resonance peaks from Figs 2(c) and 2(d) respectively. The cuts are taken with respect to the Left gate ($V_{GL}$) such that the absolute value of the current at the Apex is maximal. We now study how the splitting signature evolves as we sweep $QD_L$ through a resonance peak while keeping $QD_R$ at the positions denoted by the marks in Fig 3a,b. The percentage increase relative to the immediate background for each location relative to the Coulomb resonance is presented in Figs 2(c),2(d). We find that the highest visibility of the CPS signature occurs for the data taken at the point immediately off resonance: 96\% vs 92\% at the apex for the peak in Fig 3(a), and 71\% vs 65\% at the apex for peak in Fig 3(b). In fact, the CPS signal relative to the background remains strong though the whole resonance peak, is only reduced by a factor of 1.6 for the most off-resonance point shown.

\begin{figure}[]
\includegraphics[width=0.75 \columnwidth]{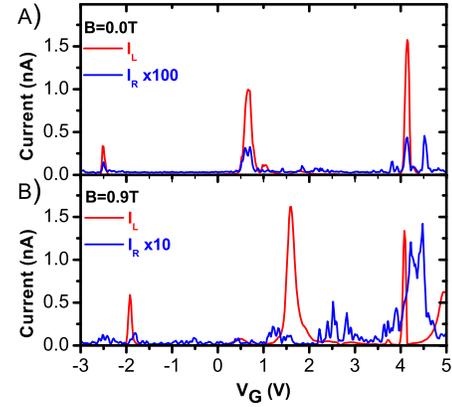}
\caption{\label{fig:overview} $I_R$ (blue) and $I_L$(red)  v.s. convolved gate voltage $V_G$ such that $QD_R$ is kept on resonance while $QD_L$ swept through several coulomb peaks. b)Data taken at zero magnetic field. Enhancement in $I_R$ can be seen when both quantum dots are on resonance. b)Data taken at $B=0.9$T (bigger than the critical field of Pb $B_C=770$mT). The background current of $QD_R$ is increased by close to a factor of $10$ due to the disappearance of the superconducting gap. Enhancement in $I_R$ no longer correlates with $QD_L$ resonances demonstrating that the feature at zero field was due to the superconducting effect, supporting Cooper pair splitting. Some areas of enhanced conductance remain; we attribute this to the change in the strength of the quantum dot tunnel barriers with respect to the change in the side gate voltages. 
}
\end{figure}

Finally, we conclusively demonstrate that the measured non-local signal truly is due to Cooper pair splitting by verifying that the signal is only measurable when Pb is in the superconducting state. We do this by applying a perpendicular magnetic field of 0.9T. This field is higher than the literature value for the critical field of lead (Pb) of $B_C\sim770$mT, and therefore reversibly eliminates superconductivity in our devices\cite{33}.  Presented in Fig 4(a) is the evolution $I_R$ while $QD_L$ is swept thought several resonances at zero magnetic field for a sample different than that in Fig 2 and Fig 3. Also in this device we confirm that  $I_R$ is enhanced when $QD_L$ is on resonance. However, when the superconducting state is broken by means of external magnetic field, no correlation between the resonances measured for $QD_R$ and $QD_L$ is apparent.

In conclusion, we achieve high efficiency Cooper pair splitting in a graphene-based, superconductor-two quantum dot junction by utilizing the unique properties of the material. The device is patterned into a true “Y-shape”, thus minimizing the separation between the quantum dot output channels. In contrast to previous devices, we induce superconductivity in the graphene via the proximity effect prior to splitting the Cooper Pairs, thus increasing the coherence length $\xi$ (while maintaining large gap $\Delta$) and ensuring that $\xi$ is larger than the QD separation $\Delta r$.  In addition, having the superconductor in plane with the quantum dots, as opposed to directly over them, we greatly increase the quantum dot charging energy $E_C$.  As a result, when both quantum dot channels are on resonance we see a Cooper pair splitting current that is up to 62\% of the total current though the device. (Much higher than previously seen in graphene\cite{18}.) Finally, by using CVD graphene, as opposed to the more common exfoliated graphene, we eliminate the need for alignment and design tailored to individual flakes, thus paving a way towards more complicated sample design and applications. 

I. V. B. acknowledges the JSPS International Research Fellowship. M. Y. and S. T. acknowledge financial support by Grant-in-Aid for Scientific Research S (No. 26220710) and Grant-in-Aid for Scientific Research A (No. 26247050). M. Y. acknowledges financial support by Grant-in-Aid for Scientific Research on Innovative Areas "Science of Atomic Layers" and Canon foundation. S. T. acknowledges financial support by MEXT project for Developing Innovation Systems and JST Strategic International Cooperative Program. S. R. and M. F. C. acknowledge financial support from EPSRC (Grant EP/J000396/1, EP/K017160, EP/K010050/1, EP/G036101/1, EP/M002438/1, EP/M001024/1), from the Royal Society Travel Exchange Grants 2012 and 2013 and from the Leverhulme Trust.


\begin{thebibliography}{99}
\bibitem{note1}Here we assume that $k_F$ is similar in both devices.
\end{thebibliography}
\end{document}